\documentclass{aa}

\begin{document}

\title{Pulsations and orbital modulation of the intermediate polar 
1WGA J1958.2+3232}

\author{ A.J. Norton\inst{1} \and H. Quaintrell\inst{1} \and
S. Katajainen\inst{2} \and H.J. Lehto\inst{2} \and
K. Mukai\inst{3} \and I. Negueruela\inst{4}}

\offprints{A.J. Norton}

\institute{Department of Physics and Astronomy, The Open University, 
	Walton Hall, Milton Keynes MK7 6AA, U.K.
 \and
	Tuorla Observatory, Turku University, V\"{a}is\"{a}l\"{a}ntie 20, 
	21500 Piikki\"{o}, Finland
\and
	NASA Goddard Space Flight Center, Laboratory for High Energy 
	Astrophysics, Code 662, Greenbelt, MD 20771, USA {\it and} Universities
	Space Research Association
\and 
	Observatoire Astronomique de Strasbourg, Rue de l'Universit\'{e} 11,
        F67000 Strasbourg, France
}

\date{Received 2001 October;
      accepted 2001 December}

\abstract{ 
We present optical photometry, spectroscopy and photopolarimetry, as
well as {\em ASCA} X-ray observations, of the recently discovered intermediate
polar 1WGA J1958.2+3232. Through the first detection of an optical beat
frequency, we confirm the previously tentative suggestion that the spin period
of the white dwarf is twice the X-ray and optical pulsation period, which we
also confirm in each case. We detect an orbital modulation in each of the U, B,
V, R and I bands for the first time, and suggest that the true orbital period is the
--1d alias of that previously suggested. We also confirm the presence of
circular polarization in this system, detecting a variable polarization which
has opposite signs in each of the B and R bands. The double peaked pulse
profile and oppositely signed polarization pulses suggest that 1WGA
J1958.2+3232 accretes onto both magnetic poles via a disc which is truncated
relatively close to the white dwarf.  
}

\maketitle

\section{Introduction}

Intermediate polars (IPs) are semi-detached interacting binaries in which a
magnetic white dwarf accretes material from a Roche-lobe filling, usually
late-type, main sequence companion star. The accretion flow from the secondary
proceeds towards the white dwarf either through an accretion disc, an accretion
stream, or some combination of both (known as disc overflow accretion), until
it reaches the magnetospheric radius. Here the material attaches to the
magnetic field lines and follows them towards the magnetic poles of the white
dwarf. The infalling material that originates from an accretion disc takes the
form of arc-shaped accretion curtains, standing above the white dwarf surface.
At some distance from this surface, the accretion flow undergoes a strong
shock, below which material settles onto the white dwarf, releasing
X-ray to optical emission. Since the magnetic axis is offset from the spin axis
of the white dwarf, this gives rise to the defining characteristic of the
class, namely X-ray (and usually optical) emission pulsed at the white dwarf
spin period. Additionally, as the X-ray `beam' sweeps around the system, there
is the possibility that some fraction of the emission will be reprocessed from
structures such as the companion star or a bulge at the edge of an accretion
disc. This will give rise to further optical emission pulsed at the lower
orbital sideband of the spin frequency, namely the spin frequency of the white
dwarf in the reference frame of the binary. Several IPs show a dominant 
optical pulsation at this orbital sideband frequency. X-ray pulsation at
the orbital sideband frequency is due to an intrinsic modulation arising from
pole-switching in the case of stream-fed accretion. Several IPs show
this type of modulation (at least at some epochs) too. About
twenty confirmed intermediate polars are now recognised with a similar number
of candidate systems having been proposed. Comprehensive reviews of various
aspects of their behaviour are given by Patterson (\cite{Patt}), Warner
(\cite{War95}), Hellier (\cite{Hell95}; \cite{Hell96}) and Norton 
(\cite{Nor95}).

\section{Observational history}

1WGA J1958.2+3232 was first identified as a $\sim 12$~min X-ray pulsator by
Israel et al (\cite{Is98}) from archival {\em ROSAT} PSPC light curves.  Israel
et al (\cite{Is99}) then identified the optical counterpart to the system as a
16th magnitude star, and claimed it was of spectral type B0Ve, indicating that
the system was a Be/X-ray binary. Negueruela et al (\cite{Neg00}) subsequently
showed that this optical counterpart had a spectrum consistent with that 
of a magnetic cataclysmic variable -- an intermediate polar. 
Uslenghi et al (\cite{Us00}) later obtained time resolved white-light optical
photometry of the star and discovered a $\sim 12$~min optical modulation with
the same period as the X-ray modulation. They also saw some evidence that the
pulse period may be twice this value. Next, Zharikov et al (\cite{Zha01})
obtained further time resolved R-band photometry and spectroscopy from which
they detected a possible orbital period of 4.36~h as well as confirming the
pulsation period as 733~sec. Most recently, Uslenghi et al (\cite{Us01})
detected circular polarization from the source in the R and I bands, with
evidence for possible modulation of the polarization at twice the previously
observed pulsation period.

\section{Observations and data reduction} 

\subsection{Optical photometry}

Time resolved photometric data were obtained at the 1~m Jacobus Kapteyn
Telescope (JKT) between 9--15 July 2000, using the SITe2 CCD and
UBVRI filters. For the majority of the observations, the CCD was windowed 
to readout a 900~$\times$~400 pixel region around the target, corresponding 
to a field-of-view of $4.5^{\prime} \times 2.0^{\prime}$. Exposures in 
different filters were interleaved each night, with a dead-time between 
exposures (for readout and filter change) of 30--40~s. Landolt standard 
fields were also observed through the relevant filters on each night. 
The observing log is shown in Table 1.

\begin{table}
\caption{Observation log for JKT photometry}
\begin{tabular}{llll} 
Night	& Filters & Exposures/s & No. of frames \\ \hline
	& & & \\
9 July	& U & 600 & 5\\
2000	& B & 60  & 7\\
	& V & 60  & 6\\
	& R & 60  & 7\\
	& I & 60  & 7\\
	& & & \\
10 July	& B & 60   & 64\\
2000	& V & 60   & 10\\
	& V & 30   & 55\\
	& & & \\
11 July	& V & 30    & 104\\
2000	& R &  30   & 101\\
	& & & \\
12 July	& V &  30   & 98\\
2000	& I &  30   & 97\\
	& & & \\
13 July	& B &  60   & 85\\
2000	& V &  30   & 85\\
	& & & \\
14 July	& V &  30   & 126\\
2000	& R &  30   & 126\\
	& & & \\
15 July	& U & 600    & 4\\
2000	& U & 300    & 15 \\
	& V & 30    & 27\\
	& & & \\ \hline
\end{tabular}
\end{table}

Bias subtraction and flat-fielding were carried out using routines in 
{\sc iraf} on all target and standard frames. Aperture photometry using
further {\sc iraf} routines  was then carried out on the target 
as well as a number of near-by comparison stars, indicated in Fig.~1.
Lightcurves containing differential photometry of 1WGA J1958.2+3232 with 
respect to each of the comparison stars were constructed. Extinction 
coefficients were then derived for each filter used on each night by 
examining the variation in observed magnitude of the comparison stars 
as a function of airmass. By combining these with the observations of
Landolt standards, this in turn allowed the derivation of zero points in
each filter for each night. Finally, these corrections were applied to the
lightcurves of the target, resulting in the calibrated lightcurves shown 
in Fig.~2.

\subsection{Optical photo-polarimetry}

 Time resolved B- and R-band circular photopolarimetry of 1WGA J1958.2+3232
 was obtained at the 2.56m Nordic Optical Telescope (NOT)
 between the nights of 28 February and 2 March, 2001. The
 Andalucia Faint Object Spectrograph and Camera (ALFOSC) was used with a
 1/4-waveplate and a calcite block. The o and e images of the star were
 separated by 15 arc sec such that the images of 1WGA J1958.2+3232 do not
 overlap with other stars in the field.

 The R-band data of 1WGA J1958.2+3232 were obtained on 1 March 2001 between
 UT 5.94 and UT 6.81, covering 2.14 spin cycles whilst the B-band data were
 obtained on 3 March 2001 between UT 5.86 and UT 6.78, covering 2.25 spin
 cycles. The seeing was between 1 and 1.5 arc seconds on both nights and
 exposure times were 58 sec for the B-band and 45 sec for the R-band
 images. On each night all the images were taken as one long sequence of
 exposures, giving a time-resolution of less than one minute.

 The CCD-frames were bias-subtracted and flat-fielded with standard
 {\sc iraf}-routines and aperture photometry was carried out with 
{\sc daophot} (Stetson
 1992). Measurements were also made for some other bright stars in the
 field of 1WGA J1958.2+3232 to calibrate the zero-point of the circular
 polarization. The sign of the circular polarization was checked by observing
 polarization standards ${\rm Grw +70^{\circ}8247}$ and LP 790--29.
 The instrumental polarization was found to be slightly dependent on the
 position of the star in the CCD-image. Corrections (1.2\% in the B-band, and
 0.9\% in the R-band) for instrumental polarization were made based on
 measurements of unpolarized field stars near the position of the object in
 the CCD-frame. 

\subsection{Optical spectroscopy}

Two spectra of 1WGA J1958.2+3232, each spanning the wavelength range $\sim
3600-6700$\AA, were obtained at the 2.5m Isaac Newton Telescope (INT) on 21
July 2000 using the Intermediate Dispersion Spectrograph, 500mm camera, R150V
grating and TEK5 CCD. This combination yields a dispersion of 3.17\AA \
pixel$^{-1}$. The exposure time for each spectrum was 1800s and the two
exposures began at 02:02UT and 02:34UT respectively. The resulting frames
were bias subtracted then flat-fielded using tungsten lamp flats; the spectra
were optimally extracted and calibrated using CuAr arcs. Fig.~3 shows the
average spectrum obtained, with the continuum normalised to unity, and emission
lines identified.

The average INT spectrum shows a similar range of emission lines to that 
reported by Negueruela et al (\cite{Neg00}). The Balmer series lines up to 
at least H$\zeta$ are all in emission, as are lines due to HeI (4472\AA,
4922\AA, 5876\AA) and HeII (4686\AA, 5412\AA). Unlike the spectra reported
by Negueruela et al (\cite{Neg00}) however, the lines are {\em not} double 
peaked.

\subsection{X-ray observations}

We have extracted the {\em ASCA} (Tanaka, Inoue \& Holt 1994) data of
1WGA J1958.2+3232 from the HEASARC.  The observation was carried
out on 1998 May 15 over an elapsed time of about 15.5 hrs.  Data
were taken with all 4 co-aligned instruments, two GIS's and two
SIS's, all in standard configuration.  We have applied the standard
screening criteria (excluding Earth occultation, passages through
the South Atlantic Anomaly, and times of high background), resulting
in $\sim$30 ksec of good data in each instrument.  We have extracted
light curves in 16s bins from each instrument in low (0.4--2 keV for
SIS, 0.7--2 keV for GIS) and high (2--10 keV) energy bands, both
from the source and from nearby source-free background region.
We have scaled the latter using the detector areas, and subtracted
from the former, to create net light curves; we have then summed
the net light curves in the low and the high bands, and also created
a total (0.4/0.7--10 keV) light curve.

\section{Results}

The average magnitudes and colours of 1WGA J1958.2+3232 derived from our seven
nights of JKT data are listed in Table 2. formal uncertainties are dominated by
the intrinsic variability of the source. We note that these magnitudes are
consistent with the UBV values reported by Uslenghi et al (\cite{Us00}) and
obtained about 1 year before our data.

\begin{table}
\caption{Average magnitudes and colours of 1WGA1958.2+3232 from JKT data}
\begin{tabular}{lr}
Filter & Magnitude \\ \hline
	& \\ 
U--V	&  $-0.4 \pm 0.1$ \\
B--V	& $0.25 \pm 0.08$ \\
V	& $15.76 \pm 0.09$\\	 
V--R	& $0.31 \pm 0.07$\\
V--I	& $0.6 \pm 0.1$\\ 
	& \\ \hline
\end{tabular}
\end{table}

\subsection{Optical power spectra}

The JKT lightcurves in magnitudes were converted into flux units prior to time
series analysis.  Fig.~4 shows the power spectrum of the longest dataset, 
i.e. the V-band lightcurve, in two panels containing the low and high 
frequency regimes. Frequencies corresponding to the peaks were measured by 
fitting Gaussian profiles to the relevant portions of the power spectra.

The power spectrum is dominated by three sets of signals: one set centred
around a frequency of $\sim 117.8$~d$^{-1}$; one set around 
$\sim 55.5$~d$^{-1}$; and a final set around $\sim 5.5$~d$^{-1}$. Each set
of peaks consists of a series of sidebands due to daily aliases either side
of a central strong peak. Clearly 
these three sets of frequencies are related to the pulse period, 
roughly twice that period, and the orbital period of the system as 
previously detected. There is also the possibility that one of the short
periods is related to a beat between the white dwarf spin and system orbital 
periods. However, it is important to note that the strongest
peaks in each of the three sets are {\em not} harmonically related to 
each other. A solution to the identification of the frequencies is as follows.

\begin{itemize}
\item We identify the strongest peak in the highest set of frequencies with 
{\em twice} the spin frequency of the white dwarf, i.e. $2\omega = 117.819 \pm
0.005$~d$^{-1}$ or $P_{\rm spin} / 2 = 733.33 \pm 0.03$~s. This is identical
with the pulse period previously detected by Israel et al (\cite{Is98}) and
Uslenghi et al (\cite{Us00}).

\item We identify one of the peaks in the middle set of frequencies as 
the beat frequency between the white dwarf spin and system orbital 
frequencies. However, exactly which of the peaks is to be identified 
with the beat frequency is not immediately clear. 

\item We identify one of the peaks in the lowest set of frequencies
with the orbital frequency of the system. However, as above, exactly which 
of the peaks is to be identified with the orbital frequency is not
immediately clear.
\end{itemize}

The strongest peak in lowest set of frequencies corresponds roughly with
the orbital period claimed by Zharikov et al (\cite{Zha01}), i.e.
$\Omega \sim  5.5$~d$^{-1}$ or $P_{\rm orb} \sim 4.4$~h. However, if this 
is chosen, it implies that the beat frequency is the --2 day alias of the 
strongest peak in the middle set of frequencies, and this peak clearly has 
a very low significance since there are many other peaks of comparable 
strength in the spectrum. Conversely, if the strongest peak of the middle 
set is the beat frequency, it implies that the orbital frequency 
is the --2 day alias of the strongest peak in the lowest set. This peak
too has a relatively low significance.  Given the relative weakness of the 
--2 day aliases of the strongest peaks in each of these two sets of 
frequencies, the most likely combination of peaks to represent the system 
frequencies is as follows.

\begin{itemize}
\item The --1 day alias of the strongest peak in the lowest set of frequencies 
is the orbital frequency of the white dwarf, $\Omega = 4.455 \pm
0.005$~d$^{-1}$ or $P_{\rm orb} = 5.387 \pm 0.006$~h. This is the --1 day
alias of the orbital period previously detected by Zharikov et al 
(\cite{Zha01}).

\item The --1 day alias of the strongest peak in the middle set of frequencies 
is the beat frequency between the white dwarf spin and the orbit, 
$(\omega - \Omega) = 54.434 \pm 0.005$~d$^{-1}$ or $P_{\rm beat} = 
1587.24 \pm 0.14$~s. 
\end{itemize}

More extreme aliases combinations are probably unlikely,
since the power at these aliases is low, although such combinations are 
possible in principle.              

Fig.~5 shows the JKT power spectra for the low 
frequency regime in each of the U, B, V, R and I bands.  The orbital 
frequency detected in the V-band (indicated by the dotted line) is also seen 
clearly in the R and B-band power spectra. Broad peaks in the I and U-band 
power spectra are consistent with the same period, possibly shifted to 
higher frequencies, but this apparent shift is probably an artifact of the 
small amount of data in these bands. We therefore detect the orbital period 
of the system in all five wave-bands.

Fig.~6 shows the same JKT power spectra, but this time in the higher
frequency regime, for all five wave-bands. Both the B and I-band lightcurves
contain signals consistent with twice the spin frequency (i.e. $2\omega$)
seen in the V band, indicated by the right-most dotted line. In the U band, 
broad peaks appear either side of this location, whilst in the R band, no 
signal is apparent. The two left-most dotted lines indicate the proposed spin
frequency ($\omega$) and the beat frequency ($\omega - \Omega$).
The beat frequency is not clearly detected in any waveband other than V, whilst 
the spin frequency itself may be detected in the R band, and possibly in the 
B band.

\subsection{Ephemerides}

From the phasing of the peaks in the Fourier transform of the JKT V-band 
lightcurve, we are able to determine spin and orbital ephemerides, based 
on our observations of:

Half the spin period:
\begin{equation}
{\rm HJD} = 2451739.318445(2) + 0.0084876(4) {\rm E}
\end{equation}

Orbital period:
\begin{equation}
{\rm HJD} = 2451739.36435(2) + 0.2245(3) {\rm E}
\end{equation}

where the zero phase is at minimum light in each case.  

We can also combine our determination of the pulse period with
earlier measurements to estimate the spin down rate of the white dwarf.
The most accurate previous determination of the pulse period is that by
Uslenghi et al (\cite{Us00}) who measured it as $733.24 \pm 0.02$s in 
June 1999. Combined with our determination of $733.33 \pm 0.03$s from
July 2000, this yields a rate of change for the period of:

\begin{equation}
\dot{P} = (2.6 \pm 1.2) \times 10^{-9}
\end{equation}

Of the other confirmed intermediate polars, PQ~Gem, V1223~Sgr and AE~Aqr are
also spinning down, but their $\dot{P}$ values are at least two orders of
magnitude smaller than this (e.g. Patterson \cite{Patt}; see also {\tt
http://lheawww.gsfc.nasa.gov/users/mukai/iphome} {\tt /iphome.html}). 
Similarly, the systems which are spinning up (BG~CMi, DQ~Her, EX~Hya, AO~Psc,
GK~Per) have $\dot{P}$ values which are also of the order $10^{-11}$. FO~Aqr
has been seen to change from spin down to spin up over the last several years,
but again the rate of period change is much smaller than the apparent change in
1WGA J1958.2+3232. Indeed the inferred $\dot{P}/P$ for 1WGA J1958.2+3232 is of
the order of $10^{-4}$~yr$^{-1}$, which is what is expected for a neutron star
accretor rather than a white dwarf (e.g. King \& Williams \cite{KW83}). It is
likely therefore that the measured pulse periods for 1WGA J1958.2+3232 are not
as accurate as claimed.

\subsection{Optical modulation profiles}

The JKT U, B, V, R and I lightcurves were folded, using the ephemerides above, 
at half the proposed spin period, at the proposed beat period, and at the
proposed orbital period of the system. After folding and binning, fluxes were
converted back to magnitudes, yielding the profiles shown in Fig.~7. The
orbital modulation is clearest in the V-band, but a similar profile with
similar phasing is also apparent in the B and R-bands. The orbital profiles in
the U and I-bands are more sparse but also consistent with the V-band profile. 
The modulation at half the proposed white dwarf spin period is clearest in the 
B and V-bands. The other bands are too noisy to draw any conclusions. Although 
an apparent modulation is seen in the I-band, it is out of phase with those 
in the B and V-bands. Of the folds at the beat period, only those in the B 
and V bands are convincing, although the profile in the R band is also 
reasonable. Much of the noise evident in all the U and I band folded profiles
is due to the relatively small amount of data in these bands.

\subsection{Optical polarimetry}

The extent of the polarimetric lightcurves are not sufficient to allow any
period searching. Instead, the reduced B- and R-band circular polarization data
and light curves were averaged into 12 phase bins over one spin cycle ($ P_{\rm
spin}=1466$ sec), according to the ephemeris in Section 4.2.  The light curves
shown in Figures 8a and 9a represent differential magnitudes with respect to
the bright field star H (see Figure 1). We note that these B and R band 
pulse profiles appear to show a single peak when folded on the 
inferred spin period, rather than the double peaked pulse profile that
would be expected from the photometry reported previously and elsewhere
in this paper. However, since the data span only just over two spin
cycles, the profiles may not be as representative as other more extensive
data sets.

The phase binned B-band curves (Figure
8) show that the circular polarization is about +0.6\% between spin phases 0.3
and 0.8, contemporaneous with a 0.1 magnitude fading, whilst the rest of
the time the polarization is close to zero. The mean circular polarization is
$+0.32\% \pm 0.10\%$, a clear $3\sigma$ detection.  In the R-band, the circular
polarization is negative (Figure 9b) with a mean level of $-1.99\% \pm 0.11\%$.
At phase 0.0 the level of polarization is around --1\% and the level increases
to --5\% (near spin phase 0.4). Just after phase 0.5 the circular polarization
drops quickly to the --1\% level. 

We also note that the circular polarization light curves of the flux
comparison star H are consistent with zero in both wavebands, varying typically
between $\pm 0.1\%$ in both B and R.

\subsection{X-ray pulsation}

Figure 10 shows the power spectrum of the {\em ASCA} X-ray lightcurve.  A
strong signal is clearly detected at a frequency of $2\omega$, but no other
harmonics or frequencies related to either the orbital or beat frequencies of
1WGA J1958.2+332 are seen. The {\em ASCA} data in two energy bands folded at
both the pulse period and twice the pulse period (i.e. the spin period of the
white dwarf) are shown in Figures 11 and 12. There is only a slight difference
between the two pulses apparent in Figure 12, as might be expected from the 
lack of a sub-harmonic in the power spectrum. 1WGA J1958.2+3232 appears to be 
a classic 'double-peaked' pulse profile intermediate polar. The modulation
depth is close to 100\% in the soft energy band, and is less in the hard
band, indicating that photoelectric absorption is a major contributor to
the overall profile. There are indications of some structure to the pulse
profiles, beyond a simple sinusoidal modulation. For instance, the `notches' 
apparent near the pulse peaks are reminiscent of those seen in AO Psc and
V1223 Sgr (Norton 1995). However, in this case they are only marginally 
significant given the error bars shown in the profiles.

\section{Discussion and Summary}

Optical photometry from the JKT has allowed us to confirm the detection of an
optical pulsation in 1WGA J1958.2+3232 at a period of 733.33~s and define a
refined ephemeris. {\em ASCA} observations also confirm the presence of an
X-ray pulsation at the same period as seen in the optical. Our detection of a
second optical pulsation period of 1587.24~s, identified with the beat period
of the system, implies that the spin period of the white dwarf is {\em twice}
the short pulsation period, i.e.  $P_{\rm spin} = 1466.66$~s and that the
orbital period is the --1 day alias of the strongest low frequency peak in our
power spectrum, i.e. $P_{\rm orb} = 5.387$~h. This is also the --1 day alias of
the previously reported orbital period. We detect an orbital modulation in each
of the U, B, V, R and I bands for the first time.

We also confirm the presence of circularly polarized emission from this source,
which is thus only the fifth IP to exhibit such behaviour. The circular
polarization is negative in the R-band and positive in the B-band, and in each
case shows evidence for variation across the spin cycle. The level of
polarization in the R-band, at $\sim -1\%$ to $\sim -5\%$, is greater
than  that seen by Uslenghi et al (2001) who saw a mean level of $\sim -0.5\%$,
whilst our detection of positive circular polarization in the B-band is a first
detection in that band.

Since the spin period of the white dwarf is twice the pulsation period observed
in both optical and X-ray flux, the pulse profile is actually double peaked. 
Following Norton et al (\cite{Nor99}), a double peaked X-ray pulse profile
indicates that two-pole accretion in 1WGA J1958.2+3232 conspires to produce two
peaks per revolution of the white dwarf. In this model, the white dwarf has a
relatively low magnetic field strength, and consequently an accretion disc
which is truncated only relatively close to the white dwarf. This leads to
large footprints of the accretion curtains and an optical depth to X-rays
across the accretion regions which is largest in a direction parallel to the
white dwarf surface, and smallest perpendicular to the surface. (This is the
reverse of the standard accretion curtain model which leads to a single peaked
X-ray pulse profile.) Confirmation of this model would require pulse phase
resolved optical spectroscopy in order to measure the direction of flow of the
accreting material corresponding to the phases of pulse maximum and minimum. 
Since this model also predicts that the accretion flow is essentially via a
disc, an X-ray beat modulation would not be expected to be present, in accord
with all observations so far, including the {\em ASCA} data presented here.

The different polarities of the circular polarization observed in the B- and
R-band data, together with light curve variations, also suggest two pole
accretion. We suggest that between spin phases 0.0 and 0.5 the negative
polarization accretion pole is seen, and after phase 0.5 emission from this
pole reduces. Between spin phase 0.3 and 0.8, the positive polarization
accretion pole is seen in the B-band. The different colours of the two
accretion poles may indicate that the poles have significantly different
magnetic field strengths, possibly as a result of an offset dipole field
structure. We note however, that the detection of circular polarization from an
IP is generally taken to be an indication of a relatively strong magnetic
field, in contradiction to the implication of the double-peaked pulse profile
model.  There is though no evidence for a strong magnetic field in this system.
The figure of $B = 8$~MG stated by Uslenghi et al (2001) is merely a value
chosen for illustrative purposes and is in no way `derived' from their
polarization measurements. We suggest that the magnetic field strength of 1WGA
J1958.2+3232 is low enough to allow a small disc truncation radius, and
consequently produce a double peaked X-ray pulse profile, but high enough that
geometrical effects allow a detection of polarized emission. For instance, in a
system with a more symmetrical field pattern than is implied here, the
(positive) polarization from one pole may be effectively cancelled out by that
(negative) from the other pole, so yielding a net polarization close to zero
and hence undetectable. It may be that the only reason we detect polarization
in 1WGA J1958.2+3232 is that the two poles have sufficiently different emission
properties that a net polarization remains. In this interpretation, the
question of whether each IP exhibits polarized emission is not only dependent
upon the strength of the white dwarf magnetic field, but also on the geometry
of the system and that of its magnetic field.

\begin{acknowledgements} 
The data analysis reported in this paper was carried out using facilities
provided by PPARC, Starlink and the Open University Research Committee. 
Astronomical computing at the OU is supported under PPARC grant
PPA/G/O/2000/00037; HQ was also supported by PPARC research grant GR/L64621
during this work.  SK wishes to thank the Finnish Academy of Sciences and
Letters (Academia Scientiarum Fennica) for support. HL wishes to acknowledge
support from grants 71355 and 44011 by the Finnish Academy.  The authors thank
Ian Howarth for acquiring and reducing the optical spectra reported herein. The
Jakobus Kapteyn Telescope and Isaac Newton Telescope are part of the Isaac
Newton Group of Telescopes on the island of La Palma. They are operated on
behalf of the UK Particle Physics and Astronomy Research Council and the
Netherlands Organisation for Scientific Research, in the Spanish Observatorio
del Roque de los Muchachos of the Instituto de Astrof\'\i sica de Canarias. The
Nordic Optical Telescope is operated on the island of La Palma jointly by
Denmark, Finland, Iceland, Norway and Sweden, in the Spanish Observatorio del
Roque de los Muchachos of the Instituto de Astrof\'\i sica de Canarias.  The
polarization data presented here have been taken using ALFOSC, which is owned
by the Instituto de Astrof\'\i sica de Andalucia (IAA) and operated at the
Nordic Optical Telescope under agreement between IAA and the NBIfAFG of the
Astronomical Observatory of Copenhagen. This research has made use of data 
obtained from the High Energy Astrophysics Science Archive Research Center
(HEASARC), provided by NASA's Goddard Space Flight Center.  {\sc iraf} is
written and supported by the {\sc iraf} programming group at the National
Optical Astronomy Observatories (NOAO) in Tucson, Arizona. NOAO is operated by
the Association of Universities for Research in Astronomy (AURA), Inc., under
cooperative agreement with the National Science Foundation.  
\end{acknowledgements}

\pagebreak

\begin{figure*}
\setlength{\unitlength}{1cm}
\begin{picture}(14,14)
\put(0,0){\includegraphics{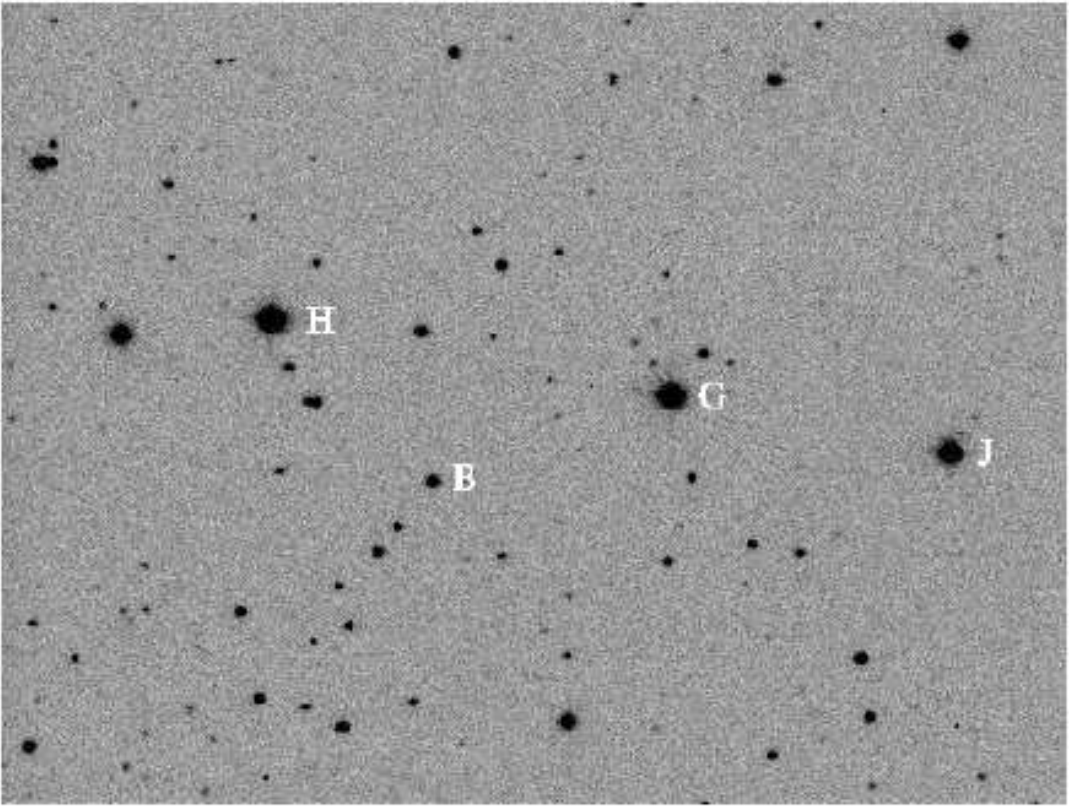}}
\end{picture}
\caption{A V-band image of the 1WGA J1958.2+3232 field covering 
$\sim 2.6^{\prime} \times 2.0^{\prime}$ obtained using the JKT. 
Following the nomenclature in Israel et al (\cite{Is99}) and Uslenghi 
et al (\cite{Us00}), the target is star B whilst stars H and J were 
used for differential photometry; star G may be variable.}
\end{figure*}

\pagebreak

\begin{figure*}
\setlength{\unitlength}{1cm}
\begin{picture}(11,21)
\put(0,0){\includegraphics{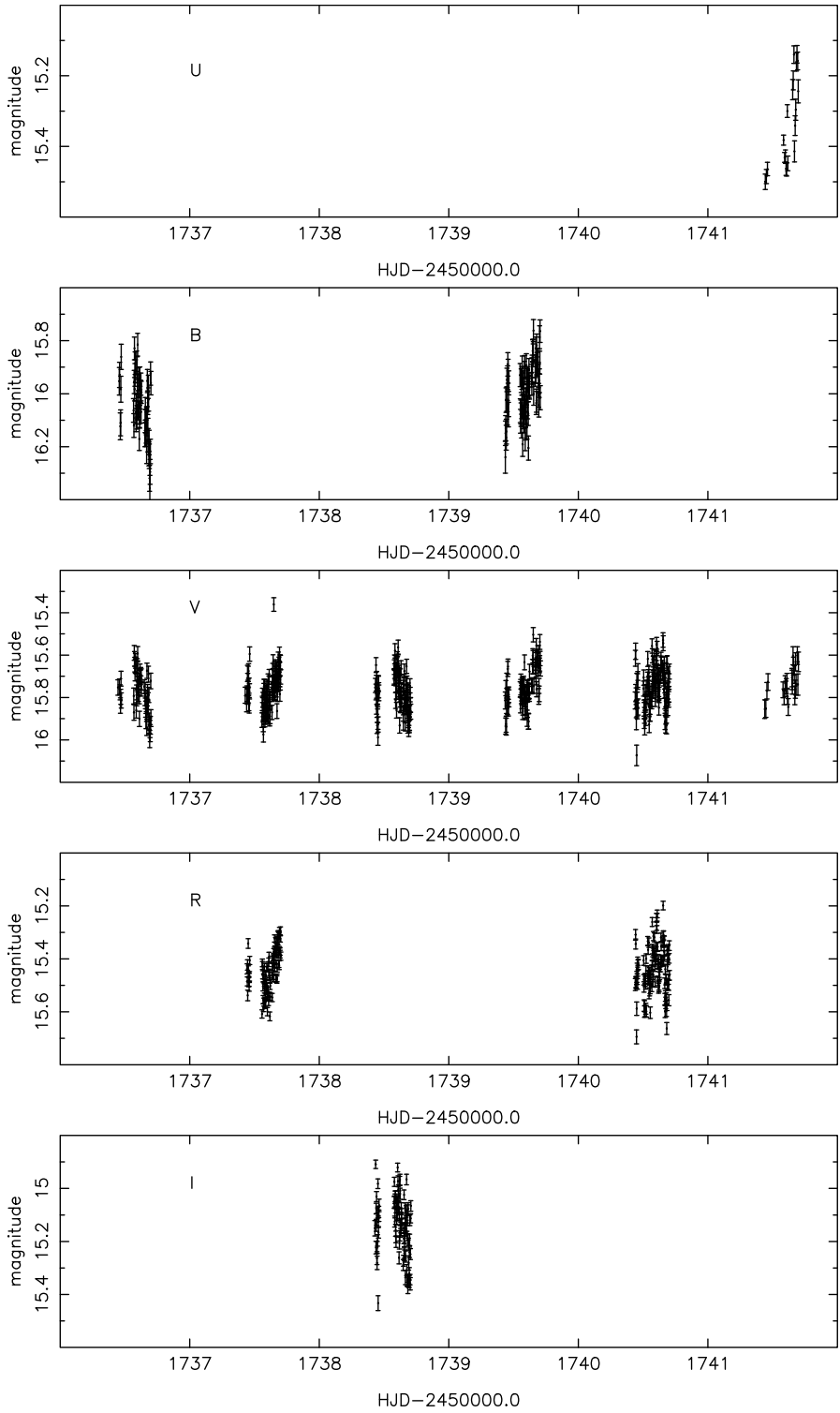}}
\end{picture}
\caption{The JKT U, B, V, R and I-band lightcurves of 1WGA J1958.2+3232 
for last six nights of the observing run.}
\end{figure*}            
        
\pagebreak    

\begin{figure*}
\setlength{\unitlength}{1cm}
\begin{picture}(14,8)
\put(-1,-2){\includegraphics{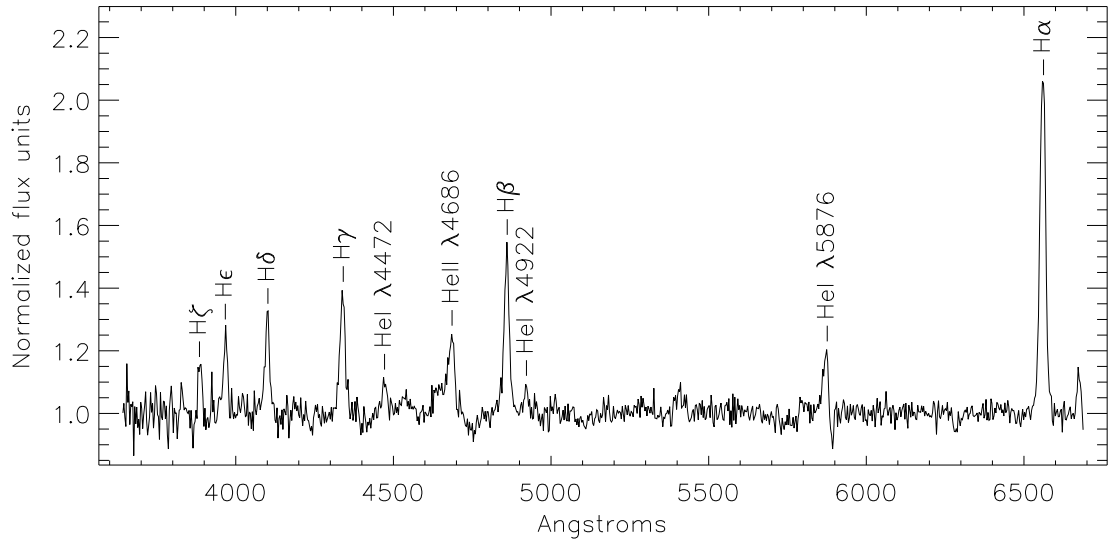}}
\end{picture}
\caption{The average INT spectrum of 1WGA J1958.2+3232 with emission lines
identified.}
\end{figure*}

\pagebreak

\begin{figure*}
\setlength{\unitlength}{1cm}
\begin{picture}(11,21)
\put(0,0){\includegraphics{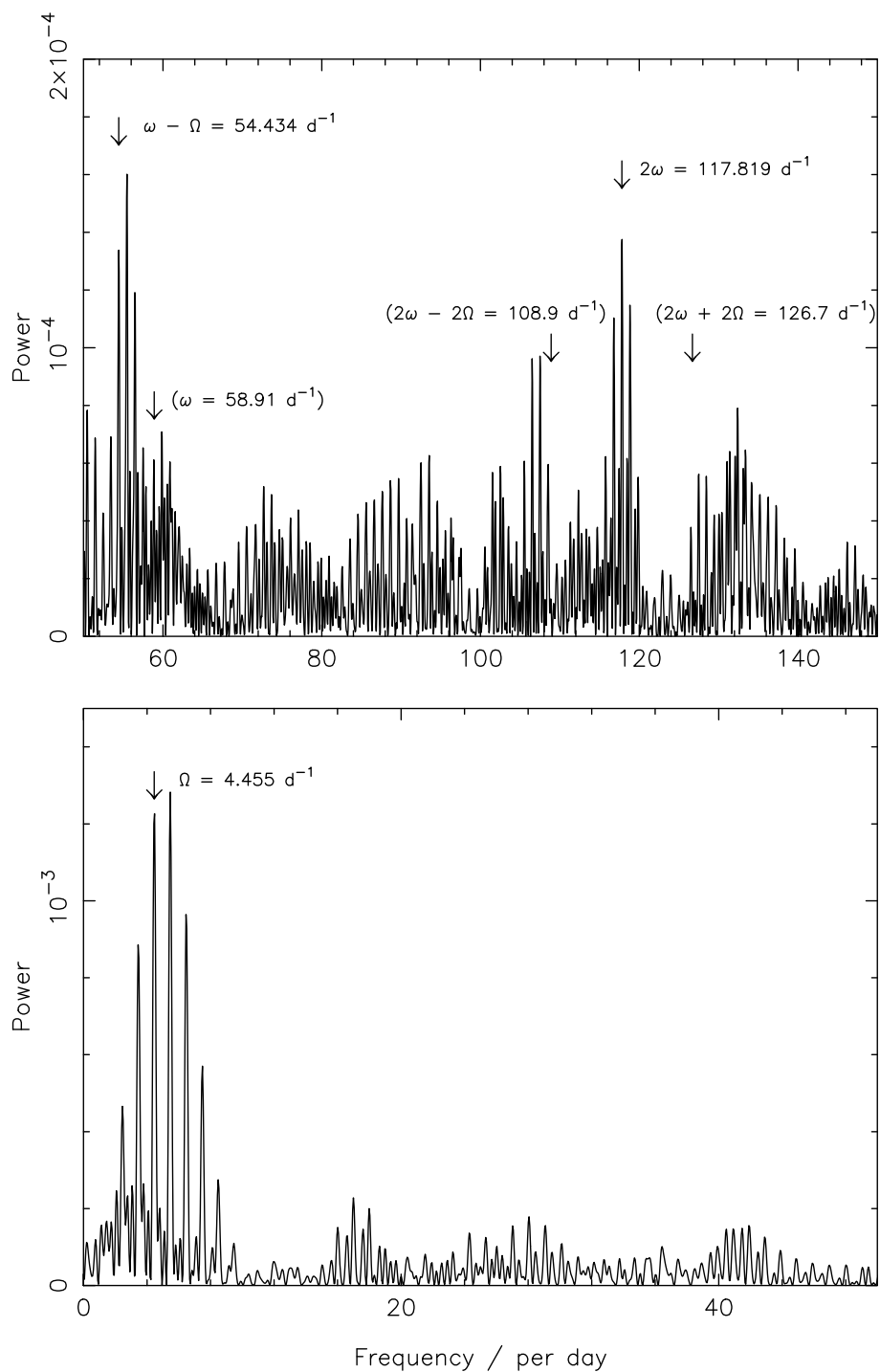}}
\end{picture}
\caption{The power spectrum of the JKT V band lightcurve of
1WGA J1958.2+3232. The upper panel shows the frequency range 50--150~d$^{-1}$,
whilst the lower panel shows the frequency range 0--50~d$^{-1}$. Note the
change in vertical scale by a factor of 7.5 between the two panels. The 
proposed spin, orbital and beat frequencies are indicated.}
\end{figure*}

\pagebreak

\begin{figure*}
\setlength{\unitlength}{1cm}
\begin{picture}(11,21)
\put(0,0){\includegraphics{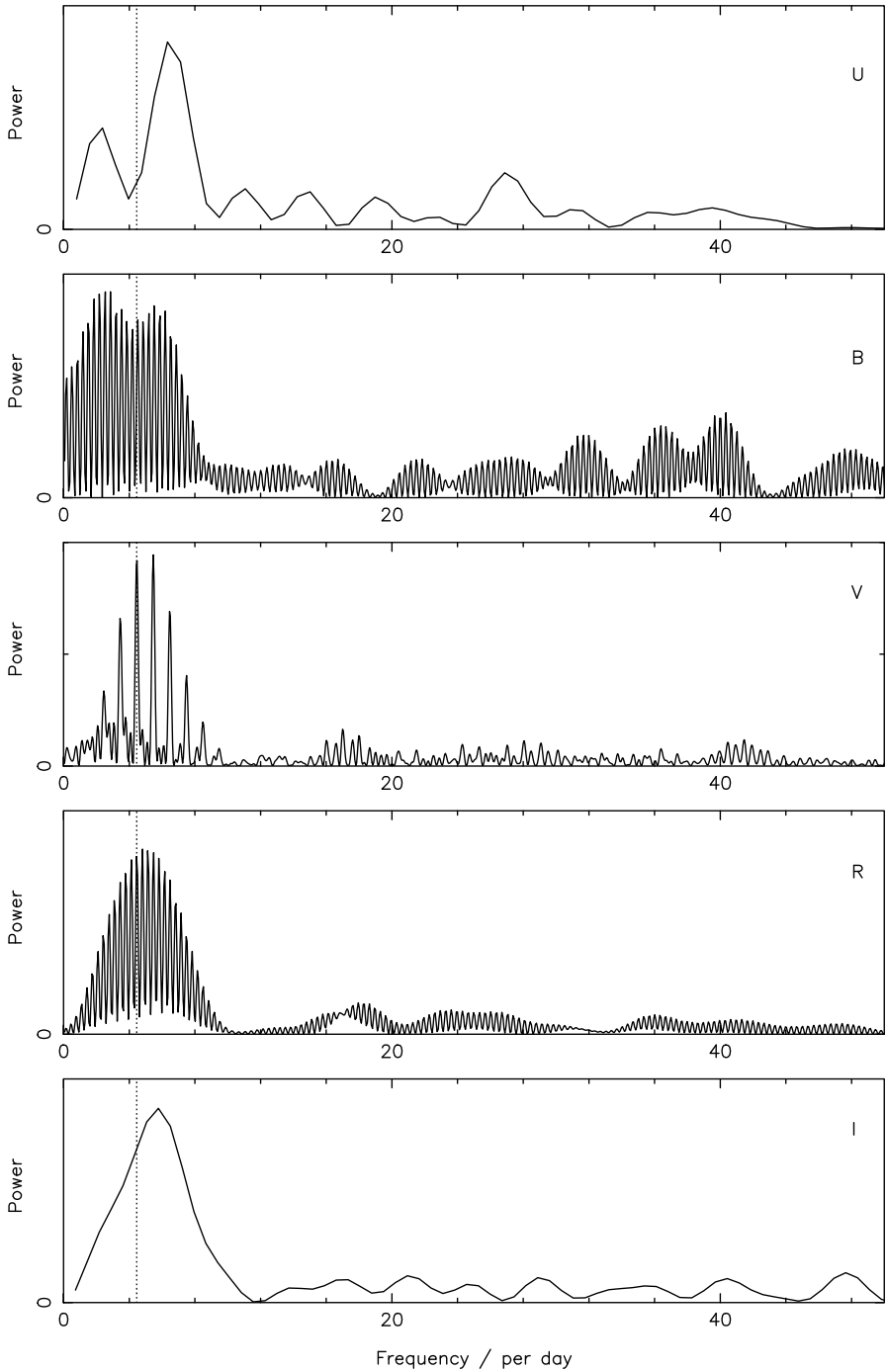}}
\end{picture}
\caption{The low frequency power spectra of the JKT U, B, V, R and I band 
lightcurves of 1WGA J1958.2+3232. The dotted line indicates the proposed 
orbital frequency ($\Omega$). Vertical axis scales are arbitrary.}
\end{figure*}            

\pagebreak

\begin{figure*}
\setlength{\unitlength}{1cm}
\begin{picture}(11,21)
\put(0,0){\includegraphics{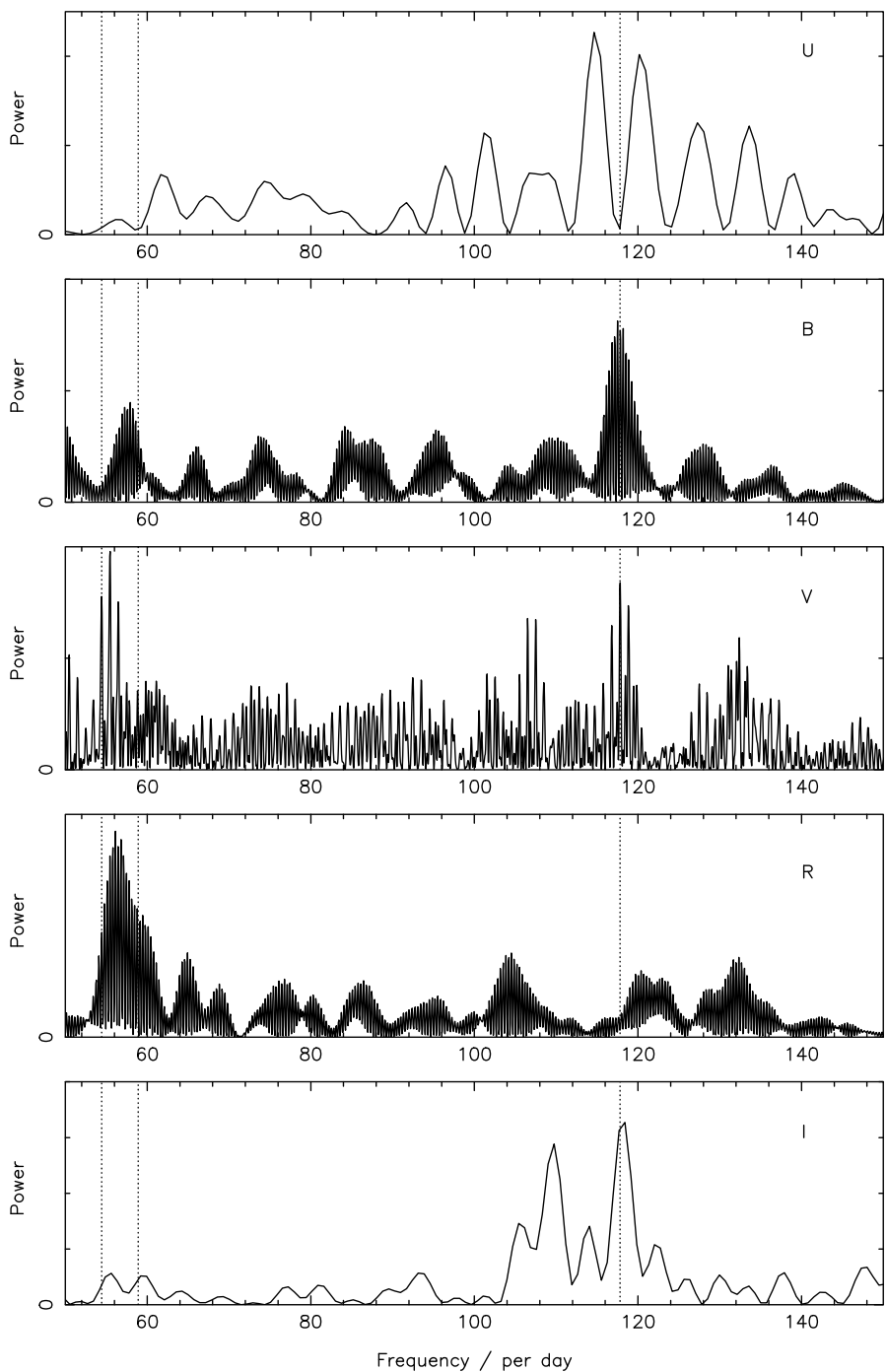}}
\end{picture}
\caption{The high frequency power spectra of the JKT U, B, V, R and I band 
lightcurves of 1WGA J1958.2+3232. Working from the right, the dotted lines 
indicate twice the proposed spin frequency ($2\omega$), the proposed spin 
frequency ($\omega$), and its orbital sideband ($\omega - \Omega$). Vertical 
axis scales are arbitrary.}
\end{figure*}

\pagebreak

\begin{figure*}[h]
\setlength{\unitlength}{1cm}
\begin{picture}(11,21)
\put(0,-1){\includegraphics{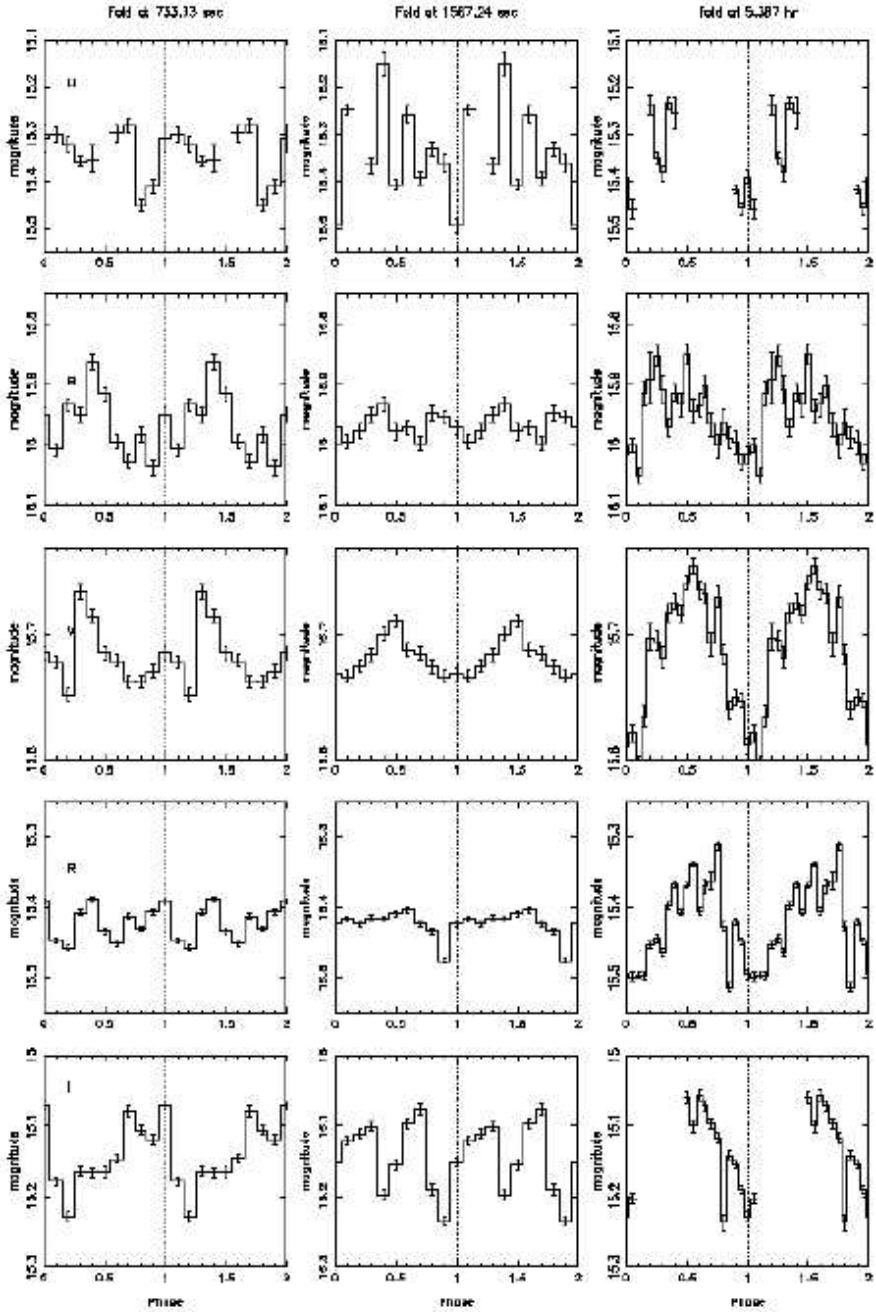}}
\end{picture}
\caption{The JKT U, B, V, R and I lightcurves of 1WGA J1958.2+3232 folded at 
half the spin period (left column), at the beat period (centre column) and 
at the orbital period (right column). Profiles are shown repeated over two 
cycles and phase zeros (minimum light) are according to the ephemerides 
given in the text. (See separate jpeg figure)}
\end{figure*}

\newpage

\begin{figure*}[h]
\setlength{\unitlength}{1cm}
\begin{picture}(14,20)
\put(0,0){\includegraphics{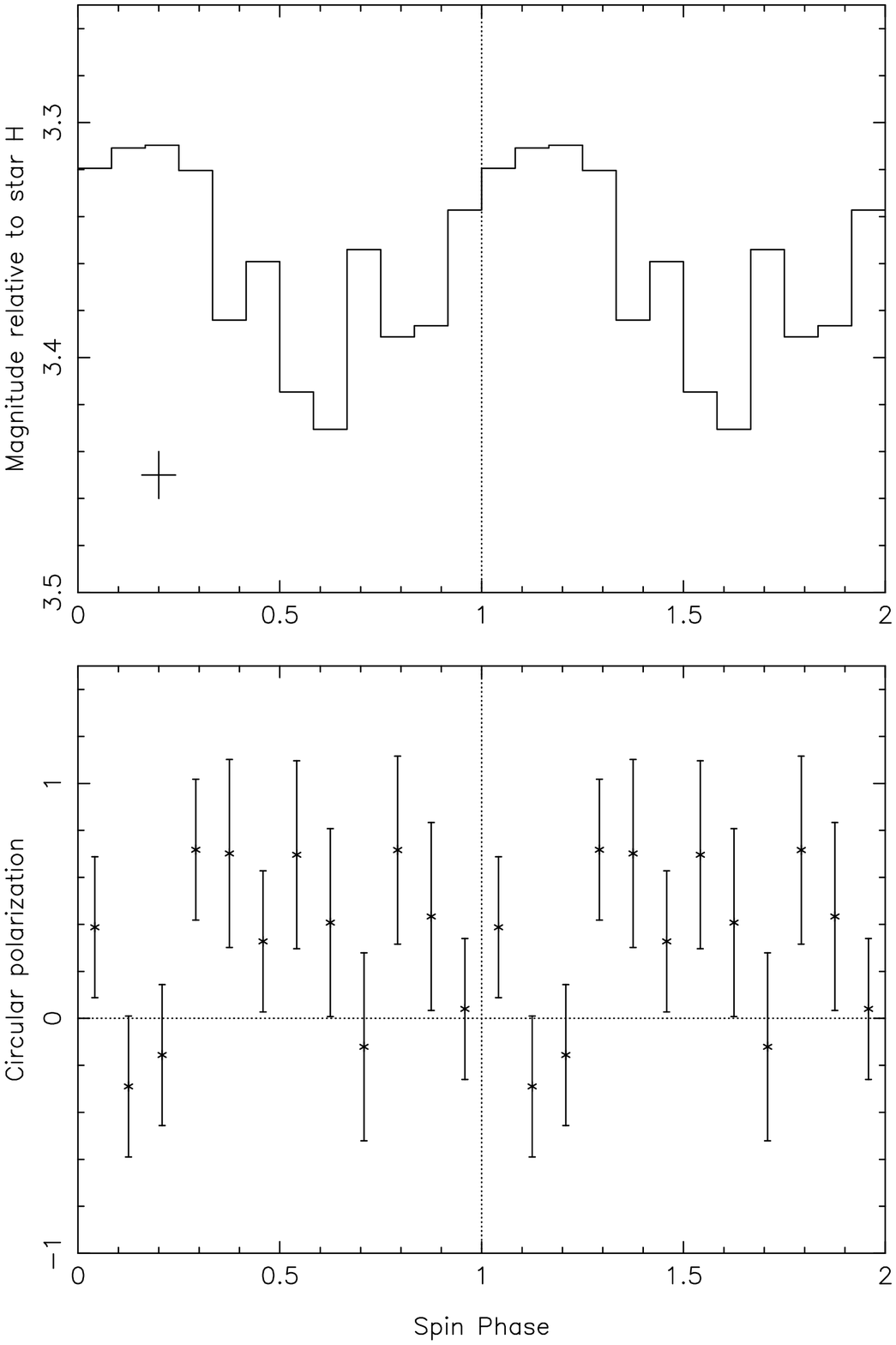}}
\end{picture}
\caption{The B-band photopolarimetry data of 1WGA J1958.2+3232 folded at 1466s
and repeated over two spin cycles.
Panel (a) shows the relative photometry with respect to star H. The typical
error bar shown is estimated from the standard deviation of the magnitude
of star H with respect to another star of constant brightness in the field.
Panel (b) shows the fractional circular polarization with one-sigma error
bars.}
\end{figure*}    

\pagebreak

\begin{figure*}[h]
\setlength{\unitlength}{1cm}
\begin{picture}(14,20)
\put(0,0){\includegraphics{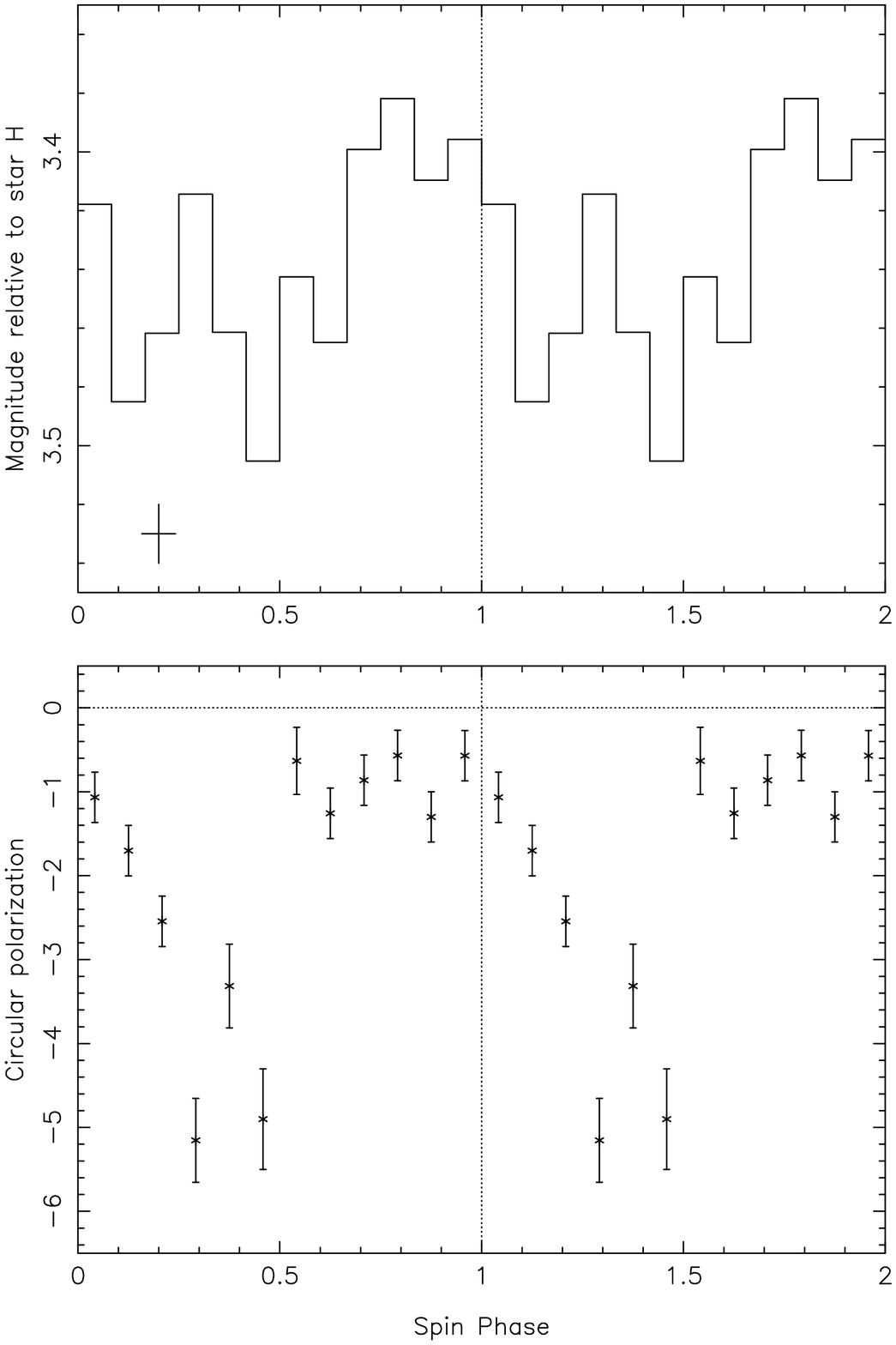}}
\end{picture}
\caption{The R-band photopolarimetry data of 1WGA J1958.2+3232 folded at 1466s
and repeated over two spin cycles.
Panel (a) shows the relative photometry with respect to star H.
The typical error bar shown is estimated from the standard deviation of the
magnitude of star H with respect to another star of constant brightness 
in the field.
Panel (b) shows the fractional circular polarization with one-sigma
error bars.}
\end{figure*} 

\pagebreak

\begin{figure*}[h]
\setlength{\unitlength}{1cm}
\begin{picture}(14,20)
\put(0,-1){\includegraphics{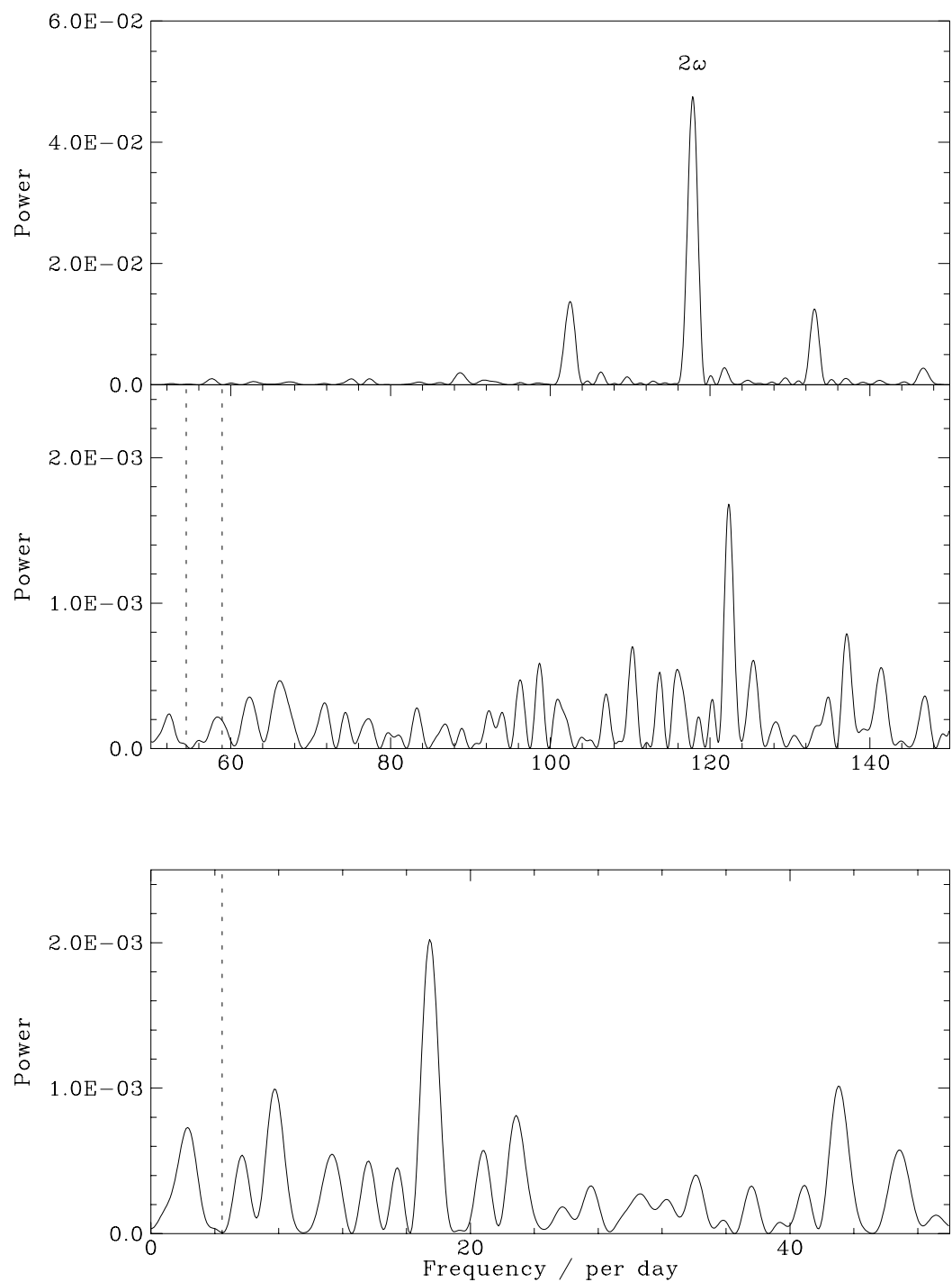}}
\end{picture}
\caption{The power spectrum of the total (0.4/0.7--10 keV) ASCA light curve
of 1WGA J1958.2+3232.  The top panel shows original power spectrum
in the frequency range 50--150d$^{-1}$. The dominant signal is at
a frequency of $2\omega$ with sidebands due to the spacecraft
orbital frequency.  The middle panel shows
the same frequency range, after the dominant signal at 117.82 d$^{-1}$
(2$\omega$) has been removed (note the change in vertical scale). The
two dashed lines show the locations of the spin and beat frequencies derived
earlier.
The bottom panel shows the power spectrum in the 0--50 d$^{-1}$ range,
after the same pre-whitening.  The dominant signal in this panel is
found near the spacecraft orbital frequency. The dashed line here shows the 
location of the orbital frequency identified earlier.}
\end{figure*}

\pagebreak

\begin{figure*}[h]
\setlength{\unitlength}{1cm}
\begin{picture}(14,20)
\put(0,-1){\includegraphics{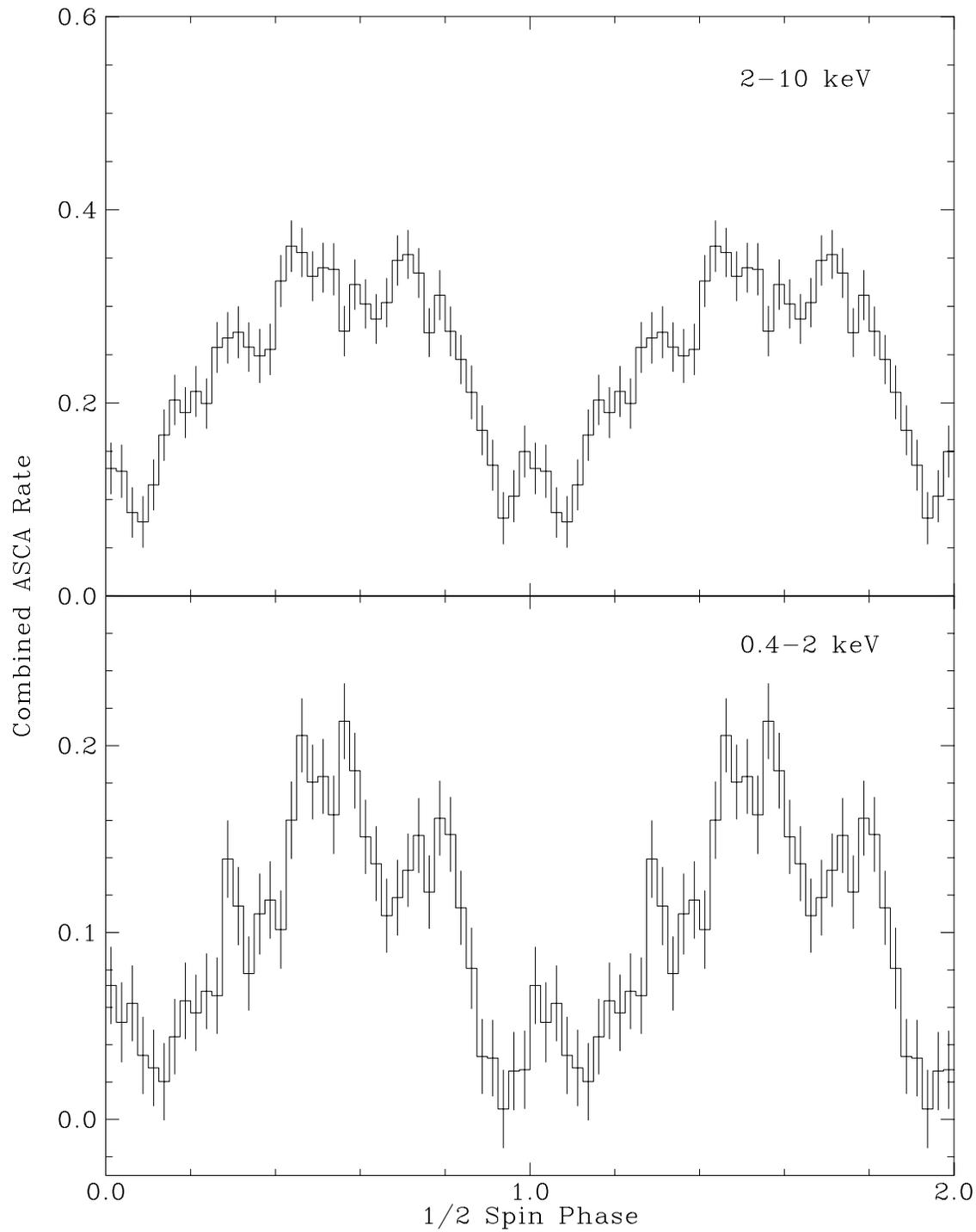}}
\end{picture}
\caption{The hard and soft {\em ASCA} lightcurves of 1WGA J1958.2+3232 
folded at the pulse period 733.33 s.}
\end{figure*}    

\pagebreak

\begin{figure*}[h]
\setlength{\unitlength}{1cm}
\begin{picture}(14,20)
\put(0,-1){\includegraphics{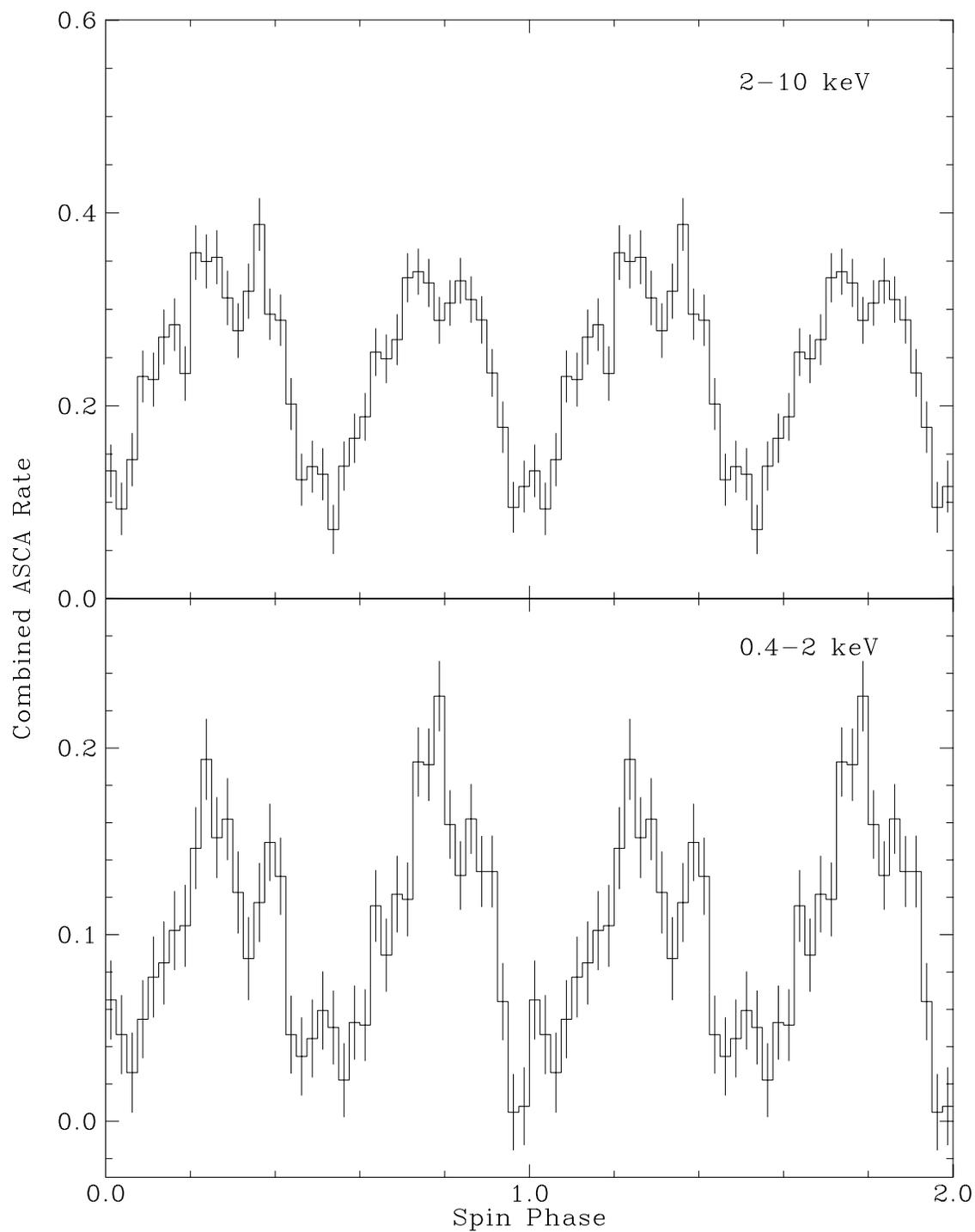}}
\end{picture}
\caption{The hard and soft {\em ASCA} lightcurve of 1WGA J1958.2+3232 
folded at the white dwarf spin period 1466.66 s.}
\end{figure*}

\end{document}